\documentclass[prd,showpacs]{revtex4}
\usepackage{amsmath}
\usepackage{graphicx}
\begin{document}
\bibliographystyle{unsrt}
\title{Casimir effect between anti-de Sitter braneworlds}
\author{James P. Norman}
\email{j.p.norman@ncl.ac.uk}
\affiliation{School of Mathematics and Statistics, University of Newcastle upon Tyne,
Newcastle upon Tyne, NE1 7RU, United Kingdom}
\date{30 June 2004}
\begin{abstract}
We calculate the one-loop effective action of a scalar field with general mass and coupling to the curvature in the detuned Randall-Sundrum brane world scenario, where
the four-dimensional branes are anti-de Sitter. We make use of conformal transformations to map the problem to one on the direct product of the hyperbolic
space $\mathrm{H}^4$ and the interval. We also include the cocycle function for this
transformation. This Casimir potential is shown to give a sizable correction to the classical
radion potential for small values of brane separation. 
\end{abstract}
\maketitle
\section{Introduction}
The one- and two- brane Randall-Sundrum models \cite{Randall:1999ee,
Randall:1999vf} have become one of the most popular
scenarios for the discussion of brane world cosmology. The two-brane model was
proposed as a solution to the the hierarchy problem without introducing
supersymmetry. 
In the original two-brane Randall-Sundrum model, the brane tensions are tuned so that the branes are flat. However, solutions also
exist when the brane tensions are detuned, with either $dS_4$ or $AdS_4$ branes
\cite{Kaloper:1999sm, Karch:2000ct}. 

The de Sitter case has been looked at
as a model for inflation \cite{Himemoto:2000nd}. 
The anti-de Sitter brane case is relevant to a supersymmetric extension
of the Randall-Sundrum model which is based on gauged five-dimensional supergravity 
\cite{Brax:2002vs,Bagger:2002rw,Bagger:2003vc}. In this model, a locally supersymmetric coupling of five dimensional gauged supergravity to branes requires
that the branes are $AdS_4$ or $M_4$. 

Upon compactification to four dimensions, the separation of the branes is promoted to a field, called the radion. In case where the branes are flat, 
the classical potential for the radion is zero. This would lead to a massless scalar particle, which is not seen in nature. This is the
radion stabilisation problem. One
possible solution is that the quantum vacuum energy of the bulk fields could generate a potential, analogous to the Casimir effect, to
stabilise of the radion. This mechanism has been looked at in \cite{Goldberger:2000dv,Garriga:2000jb,Flachi:2001pq,Hofmann:2000cj,Saharian:2002bw,Knapman:2003ey} for bulk scalar fields, and \cite{Flachi:2001bj,Flachi:2001ke}
for fermions, 
where it was found that it is not possible to simultaneously
solve the hierarchy problem and have an acceptable mass for the radion. 
A related calculation in five dimensional heterotic M theory has been performed in
\cite{Garriga:2001ar,Moss:2004un}. 

When the branes are $AdS_4$ or $dS_4$, the brane separation is fixed by the junction conditions and there is a classical
potential for the radion \cite{Bagger:2003dy}. The potential is a stabilizing one in the $AdS_4$ case, and an unstable one in the $dS_4$ case. 
The Casimir effect would then be a quantum correction to the classical potential. In the $dS_4$ brane case, the quantum effective
potential has been calculated for conformally coupled scalar fields in \cite{Naylor:2002xk} and for massless
fermions in \cite{Moss:2003zk}. Small deviations from the conformal values
were considered in \cite{Elizalde:2002dd}.

The Casimir effect between two anti-de Sitter branes has not yet received 
much attention. This
may be because the model is not as relevant to cosmology as its de Sitter
cousin. However, there seems to be no \emph{a priori} reason to favour the
$dS_4$ case over the $AdS_4$ case, since the brane tensions are put in ``by hand''.
The Casimir effect in the supersymmetric $AdS_4$ case may be important as it may
give a dynamical mechanism for 
breaking supersymmetry (see \cite{vonGersdorff:2002tj} 
for a related calculation on a flat orbifold). However, the calculation of the effective 
action of the higher spin fields in the supergravity multiplet in this background 
presents a difficult technical challenge. To simplify the calculation and illustrate our
method, here we will consider only a scalar field in
the background with $AdS_4$ branes, 
with the hope that many of the features of the calculation apply to higher spin fields.
Some related work on the one-loop effective potential 
of $AdS_n \times S^n$ was performed in \cite{Uzawa:2003ji}. Spontaneous
generation of $AdS_4$
branes in an $AdS_5$ bulk has been investigated in \cite{Nojiri:2000eb,Nojiri:2000bz}.

We calculate the one-loop effective action for scalar fields with general mass
and coupling to the scalar curvature. Both Dirichlet and Robin type boundary
conditions will be considered. It will be shown that the quantum effective potential
can give a sizable correction to the classical potential for small values of the
radion. Our calculation cannot be carried over easily to the $dS_4$ case, because
of the different global properties of $dS_4$ and $AdS_4$.

We choose to work in the ``downstairs'' picture of a manifold with boundary and
impose boundary conditions on the fields,
rather than the ``upstairs'' picture of an orbifold with singular branes. The one-loop effective action of the scalar field is evaluated 
using $\zeta$ function regularization. We use the properties of the effective
action under conformal rescalings of the metric to relate the effective action
to that of a scalar field on the direct product of $AdS_4\times I$. 
$\zeta$ function regularization requires that we rotate to the appropriate
Euclidean space, which is $\mathrm{H}^4 \times I$, where $\mathrm{H}^4$ is the four dimensional hyperbolic space. We then calculate the
correction factor for this transformation, known as the cocycle function. Our
curvature conventions are given in \cite{Moss:2004un}.

\section{Operators and Background Metric}

The background metric for the Randall-Sundrum model with $AdS_4$ branes can be written as
\begin{equation}\label{eq:metric}
ds^2=e^{-2\omega(z)}\left(g^{(4)}_{\mu \nu} dx^\mu dx^\nu + dz^2 \right)
\end{equation}
where $g^{(4)}_{\mu \nu}$ is the metric of $AdS_4$ with $AdS$ radius $a$, 
\begin{equation}
\omega(z)=\ln \left(\sqrt{\frac{|\Lambda|}{6}} a \sin \left( z/a\right) \right),
\end{equation}
and $\Lambda$ is
the bulk cosmological constant.
In the absence of branes, the conformal coordinate $z$ would run from $0<z<a\pi$. However,
in the ``downstairs'' picture of a manifold with boundaries, the branes cut off the space and
$z$ is restricted to the space between the branes. The positions
of the branes are fixed at the classical level by junction conditions on the
metric. We call the positions of the
branes $z_1$ and $z_2$. Explicitly, the classical positions of the branes are
\begin{equation}
z^{\text{classical}}_{1,2} = a\arccos{\frac{\sigma_{1,2}}{\sigma}},
\end{equation}
where $\sigma=\sqrt{|\Lambda|/6}$, $\sigma_1= T_1/6$, $\sigma_2= -T_2/6$, and 
$T_{1,2}$ are the
tensions of the branes at $z_1$ and $z_2$, respectively. However, these values
may receive corrections from quantum effects, so we keep $z_1$ and $z_2$
general. One condition we can always impose is
that the metric on the brane at $z_1$ has a scale factor of unity.
This fixes $\sin(z_1/a)=(\sigma a)^{-1}$. Without loss of generality, we also
restrict $z_2>z_1$.

In the limit $a\rightarrow \infty$, the function $w(z)\approx \ln (\sigma z)$, 
and we recover the flat brane Randall-Sundrum metric. In this limit, 
the positions of the branes are no longer determined by the junction 
conditions, with all brane positions being allowed.

We consider the effective action of a scalar field $\phi$ with general mass $m$
and coupling to the scalar curvature $R$. The fluctuation operator is
\begin{equation}\label{eq:operator}
\Delta=-\nabla^2+\xi R + m^2.
\end{equation}

The one-loop effective action will be regularized using $\zeta$-function regularization. We work in Euclidean space, with $AdS_4$ becoming the hyperbolic
space $\mathrm{H}^4$.
The $\zeta$ function is defined as the trace of the operator $\Delta$
to some power $-s$, in some region of the complex plane where the trace converges, i.e.,
\begin{equation}
\zeta(s)=\text{tr} \Delta^{-s}=\sum_n\int_0^\infty d \lambda\  \mu(\lambda) d_n \rho_n(\lambda)^{-s}.
\end{equation}
Here, the eigenvalues of the operator $\rho_n(\lambda)$ are assumed to split into a 
continuous part, labeled by the
real parameter $\lambda$, and a discrete part, labeled by integers $n$. 
The spectral
function $\mu(\lambda)$ gives a ``density of states'' in the continuous spectrum, and
is analogous to the discrete degeneracy factor $d_n$.
The one loop effective action is then defined to be
\begin{equation}
W=-\frac12\zeta'(0)-\frac12\zeta(0)\log \mu_R^{2},
\end{equation}
where we have analytically continued the $\zeta$ function to $s=0$. The renormalisation
scale $\mu_R$ has been introduced to make the eigenvalues dimensionless. We can also define 
the effective potential on the brane at $z_1$ by dividing
throughout by the volume of the $\mathrm{H}^4$ space. That is, we define the 
one-loop effective potential $V$ through
\begin{equation}
W=\int|g^{(4)}| d^4 x \ V.
\end{equation}

Rather than work directly with the eigenvalues of Eq.(\ref{eq:operator}), which
are difficult to obtain in curved space, we can use the behaviour of the effective
action under conformal rescalings of the metric 
to simplify the problem. Considering operators of Laplace type, with
$\Delta = -\nabla^2+X$, we introduce
a one-parameter family of metrics $g^\epsilon$ related to the physical metric
by a conformal rescaling, so that 
\begin{equation}
g_{\alpha \beta}^\epsilon={\Omega(\epsilon)}^2 g_{\alpha \beta},
\quad \Omega(\epsilon)=e^{(1-\epsilon)\omega(z)}.
\end{equation}
The conformally rescaled operator is
$\Delta_\epsilon=-\nabla_\epsilon^2+X_\epsilon$, where 
$X_\epsilon=X\Omega(\epsilon)^{-2} - \frac{3}{16} 
\left(R \Omega(\epsilon)^{-2} -R_\epsilon \right)$.
One can then show that \cite{Dowker:1990ue}
\begin{equation}
W[\epsilon=1,\Delta]=W[\epsilon=0, \Delta_0]
+C\left[\Omega \right],
\end{equation}
where the cocycle function $C\left[\Omega \right]$ is given (in five dimensions) 
in terms of
the generalized heat kernel coefficient $B_{5/2}(f,\Delta)$ as
\begin{equation}\label{eq:cocycle}
C[\Omega]=\int_{0}^{1} d\epsilon \ 
B_{5/2}\left(\omega,\Delta_\epsilon\right).
\end{equation}
Hence, we can relate the one-loop effective action of the scalar field in
the warped metric (\ref{eq:metric}) to one on the direct product manifold
$\mathrm{H}^4 \times I$. 

We will first consider
the effective action of the conformally transformed operator. The
conformally transformed operator $\Delta_0$ separates into
\begin{equation}
\Delta_0 = \Delta_{I}+\Delta^{(4)},
\end{equation}
where $\Delta_{I}$ contains all dependence on the $z$-direction
\begin{equation}\label{eq:DeltaI}
\Delta_{I} = -\partial_z^2 - \xi \frac{12}{a^2} +
\left(\xi-\frac{3}{16}\right) \left(8 \omega''-12 \omega'^2 \right) + 
m^2 e^{-2 \omega},
\end{equation}
and ${\Delta^{(4)}}$ is the 4-dimensional Laplacian for a massless
scalar field on $\mathrm{H}^4$.
The eigenvalues and $\zeta$ function for $\Delta^{(4)}$  have been
calculated by Camporesi \cite{Camporesi:1991nw}. The eigenvalues are
continuous and labeled by the real parameter $\lambda$. The eigenvalues
of $\Delta_I$, which we denote $m_n^2$, are discrete.
After performing a separation of variables, we can regard the complete $\zeta$-function as a sum over $\zeta$-functions
of scalar fields with mass $m_n$ on $\mathrm{H}^4$ \footnote{In fact, $\Delta_I$ is
the Kaluza-Klein mass operator}. Thus, from  \cite{Camporesi:1991nw}, we
find
\begin{eqnarray}\label{eq:zeta}
\zeta(s)= \frac{a^{2s-4}}{8\pi^2} \int {|g^{(4)}|}^{1/2} \sum_n &&\left\{\frac{b_n^{2-2s}}{8(s-1)}+
\frac{b_n^{4-2s}}{2(s-1)(s-2)} \right. \nonumber \\ 
&& \left. - 2\int_0^\infty d\lambda
\frac{\lambda\left(\lambda^2+\frac14\right)}
{\left(1+e^{2\pi \lambda}\right){(\lambda^2+b_n^2)}^{s}} \right\},
\end{eqnarray}
where we must restrict $s>5/2$, and we have introduced $b_n^2=\frac94 + m_n^2 a^2$. 

The eigenvalues $m_n^2$ of the operator $\Delta_{I}$ now need to be found.
We consider both Dirichlet and Robin boundary conditions on the field at the boundaries 
$z_1$ and $z_2$. However, we shall illustrate our method using Dirichlet boundary conditions,
and simply quote results for Robin boundary conditions in Appendix~\ref{sec:robin}, as
the calculation is similar. The eigenfunctions of  Eq. (\ref{eq:DeltaI})
can be written in terms of associated Legendre functions of  
order
$\mu=\sqrt{4-20\xi+ m^2/\sigma^2}$ and degree $-1/2+b_n$.  
For later convenience, we define the functions
\begin{subequations}
\begin{eqnarray}\label{eq:RandS}
R^{-\mu}_{-1/2+b_n} (\theta) &=& \sqrt{\sin\theta} 
\left(\mathrm{P}^{-\mu}_{-1/2+b_n} (\cos\theta)\right),
\\
S^{\mu}_{-1/2+b_n} (\theta) &=& \sqrt{\sin\theta}\left(
\mathrm{P}^{\mu}_{-1/2+b_n} (\cos\theta)
-\frac{2i}{\pi}\mathrm{Q}^{\mu}_{-1/2+b_n} (\cos\theta)\right).
\end{eqnarray}
\end{subequations}
These functions are linearly independent for all $\mu$ and $b_n$. The general
solution to $\Delta_I f_n = m_n^2 f_n$ is then a linear combination of $R^{-\mu}_{-1/2+b_n}(z/a)$
and $S^{\mu}_{-1/2+b_n}(z/a)$.
Applying Dirichlet boundary conditions $\phi = 0$ on the boundaries $\partial {\cal M}$,
leads to an implicit equation for $b_n$ through
\begin{equation}
F(b_n)=R^{-\mu}_{-1/2+b_n} (\theta_1)S^{\mu}_{-1/2+b_n}
(\theta_2)-R^{-\mu}_{-1/2+b_n} (\theta_2)S^{\mu}_{-1/2+b_n}(\theta_1)=0,
\end{equation}
where $\theta_{1,2}=z_{1,2}/a$.

\section{Zeta function on $\mathrm{H}^4 \times I$}
The sums over the discrete eigenvalues $b_n$ in Eq. (\ref{eq:zeta}) are complicated
by the fact that we only know the $b_n$ through the implicit equation
$F(b_n)=0$. However, we can use techniques developed for studying the
Casimir effect on balls and spheres \cite{Elizalde:1993,Bordag:1996gm} to
convert the sums into contour integrals of the function $F$. We will first consider the terms
$\sum_n b_n^{2-2s}$ and $\sum_n b_n^{4-2s}$ in the $\zeta$ function. Our procedure
follows that of the flat brane case, in that we write the sum as 
\begin{equation}
\hat{\zeta}(s)=\sum_n b_n^{-s} = \frac{1}{2\pi i} \int_{\cal C} dz z^{-s} \frac{F'(z)}{F(z)},
\end{equation}
where the contour $\cal C$ encloses all the real positive roots of $F(z)=0$. As in the
flat brane case, we wish to write this contour integral as an integral over the imaginary
axis. To do this, we need to know the asymptotic behaviour of $F(ix)$ for large $x$.
We define the functions $\Sigma^R(x,\theta)$ and $\Sigma^S(x,\theta)$ by
\begin{equation}\label{eq:SigmaR}
R^{-\mu}_{-1/2+ix}
(\theta) = e^{i\pi \mu/2+i\pi/4} \sqrt{\frac{1}{2\pi}} e^{x \theta}
\frac{\Gamma(ix-\mu+1/2)}{\Gamma(ix+1)} \Sigma^R(x,\theta) 
\end{equation}
and
\begin{equation}\label{eq:SigmaS}
S^{\mu}_{-1/2+ix}
(\theta) = e^{i\pi \mu/2-i\pi/4}\sqrt{\frac{2}{\pi}} e^{-x \theta}
\frac{\Gamma(ix+\mu+1/2)}{\Gamma(ix+1)} \Sigma^S(x,\theta).
\end{equation}
From the representation of the Legendre functions in terms of 
hypergeometric series \cite[page 146]{Bateman}, it can be seen that 
$\Sigma^R(x,\theta)$ and $\Sigma^S(x,\theta)$ have asymptotic series of the
Poincar\'e type --- that is, asymptotic expansions in
inverse powers of $x$. We can also use properties of the $\Gamma$ function
\cite[page 256]{Abramowitz} to show
\begin{equation}
\ln \left|\frac{\Gamma(ix+\mu+1/2)\Gamma(ix-\mu+1/2)}{\Gamma(ix+1)\Gamma(ix+1)}\right|
=\ln \left|\frac{\sinh{(\pi x)}}{x\cosh{\left[\pi(x-i\mu)\right]}}\right| \sim -\ln(x)+O(e^{-x}).
\end{equation}
Hence, for large $x$, 
\begin{equation}
\ln |F(ix)| \sim  (\theta_2-\theta_1)x +\chi \ln(x) -\ln \pi + O(x^{-2}),
\end{equation}
where
\begin{equation}
\chi=-1.
\end{equation}
Therefore, for $s>1$, 
the contour can be deformed to an integral over the imaginary axis. After
a few manipulations, we find
\begin{eqnarray}\label{eq:zetahatreg}
\hat{\zeta}(s)&=&\frac{\sin{\frac{\pi s}{2}}}{\pi} \int_\varepsilon^\infty dx x^{-s} \frac{d}{dx} \ln
\left|\frac{F(ix)}{F_a(ix)}\right|
+\frac{\sin{\frac{\pi s}{2}}}{\pi} \int_\varepsilon^\infty dx x^{-s} \frac{d}{dx} \ln \left|F_a(ix)\right|
\nonumber \\
&&+\frac{1}{2\pi i} \int_{{\cal C}_\varepsilon} dz z^{-s} \frac{F'(z)}{F(z)}
\end{eqnarray}
where ${\cal C}_\varepsilon$ is a small semicircle of radius $\varepsilon$ 
around the origin, and 
\begin{equation}
F_a(ix)= R^{-\mu}_{-1/2+ix} (\theta_2)S^{\mu}_{-1/2+ix}(\theta_1).
\end{equation}

The first term on the RHS of Eq. (\ref{eq:zetahatreg}) can now be continued to $s=-2$ and $s=-4$.
The second term still cannot be analytically continued to these points as the integrals
diverge at large $x$ if we take $s<1$. We therefore add and
subtract terms which cause the integral to diverge in this region
to enable us to analytically continue this term to $s=-2N$, where $N$ is a
positive integer. If we define $r_k$ and $s_k$ as the coefficients in the asymptotic
expansion of $\ln |\Sigma^R(x,\theta)|$ and $\ln |\Sigma^S(x,\theta)|$, i.e.,
\begin{equation}
\ln |\Sigma^R(x,\theta)|\sim\sum_{k=1}^\infty r_k (\theta) x^{-k}+O(e^{-x}),
\quad \ln |\Sigma^S(x,\theta)|\sim\sum_{k=1}^\infty s_k (\theta) x^{-k},
\end{equation}
then we can define
\begin{eqnarray}\label{eq:Fareg}
\ln {|F_a(ix)|}_{\text{reg}}&=&
\ln {|F_a(ix)|}+\ln \pi-(\theta_2-\theta_1)x-\chi\ln(x) \nonumber \\
&&-\sum_{k=1}^{2N-1} \left(r_k(\theta_1)+s_k(\theta_2)\right)
x^{-k} -\left(r_{2N}(\theta_2)+s_{2N}(\theta_1)\right)
x^{-2N}e^{-1/x}.
\end{eqnarray}
so that $\ln {|F_a(ix)|}_{\text{reg}} \sim  O(x^{-2N-1})$ for large $x$. We
can now substitute the for $\ln {|F_a(ix)|}$ in terms of 
$\ln {|F_a(ix)|}_{\text{reg}}$ in Eq. (\ref{eq:zetahatreg}), and analytically perform
the integrals of the ``extra'' terms in the right-hand side of Eq. (\ref{eq:Fareg}).
We can now analytically continue to $s=-2N$.
Provided that $F(x)$ is finite as $x\rightarrow 0$, the contribution from the integral around
the small semicircle ${\cal C}_\varepsilon$ vanishes when we take $\varepsilon \rightarrow 0$.
Thus, we find
\begin{equation}
\hat{\zeta}(-2N)=-(-1)^N N \left(r_{2N}(\theta_2) + s_{2N}(\theta_1) \right),
\end{equation}
and
\begin{eqnarray}
\hat{\zeta}'(-2N)&= -N(-1)^N &\left\{\int_0^\infty dx x^{2N-1} \ln
\left|\frac{F(ix)}{F_a(ix)} \right|+ \int_0^\infty dx x^{2N-1} \ln
{|F_a(ix)|}_{\text{reg}}\right.\nonumber \\ 
&&\left.-\left(\gamma+\frac{1}{2N}\right)\left(r_{2N}(\theta_1)+s_{2N}(\theta_2)\right)\right\}.
\end{eqnarray}
where $\gamma$ is Euler's constant.
We will quote the value of the first four coefficients $r_k$ and $s_k$ in Appendix~\ref{sec:coeff}.

For the third term in Eq. (\ref{eq:zeta}), we can interchange the order of 
integration and summation for large $s$. The resulting sum is not of the form considered above, 
but is a generalized Epstein type $\zeta$ function. Following Bordag \emph{et al.} \cite{Bordag:1996gm}, 
we can again write the sum as a contour integral of $F$
around a contour which encloses all the positive zeros of $F$.
\begin{equation}\label{eq:zetaE}
\tilde{\zeta}(s)=\sum_n \left(\lambda^2+b_n^2\right)^{-s}
= \frac{1}{2\pi i} \int_{\cal C} dz \left(\lambda^2+z^2\right)^{-s} \frac{F'(z)}{F(z)}.
\end{equation}

For $1/2<s<1$, we can proceed in a similar manner to the above
and deform the contour to an integral over the imaginary axis. This gives
\begin{equation}\label{eq:zetatildereg}
\tilde{\zeta}(s)=\frac{\sin \pi s}{\pi} \int_{\lambda}^\infty dx \left(x^2-\lambda^2\right)^{-s} \frac{d}{dx} \ln |F(ix)|.
\end{equation}
We again add and subtract the terms that cause the integral to diverge at
$s=0$. We find that the analytic continuation of Eq. (\ref{eq:zetaE}) at $s=0$ is
\begin{eqnarray}
\tilde{\zeta}(0) &=& \frac{\chi}{2}, \label{eq:epszeta0}\\
\tilde{\zeta}'(0)&=&-\ln |\pi F(i\lambda)| \label{eq:epszetaprime0}	.   
\end{eqnarray}
Hence, we can now write down the complete $\zeta$ function 
and its derivative at $s=0$. We find
\begin{eqnarray}
\zeta(0)&=&\frac{1}{16\pi^2 a^4}\int|g^{(4)}| d^4 x
\left\{-\chi \frac{17}{960} - \sum_{i=1,2}\left(r_4(\theta_i)+\frac14 r_2
(\theta_i) \right)\right\}, \label{eq:zeta0}\\
\zeta'(0)&=&-\frac{a^{-4}}{8\pi^2}\int|g^{(4)}| d^4 x 
\left(G(\theta_1,\theta_2) + C_1(\theta_1)
+ C_2(\theta_2)+q\right) \nonumber \\
&&+\ln a^2 \zeta(0) - 2 \gamma \zeta(0),
\label{eq:zetaprime0}
\end{eqnarray}
where the ``non-local'' part in Eq. (\ref{eq:zetaprime0}) is
\begin{eqnarray}\label{eq:nonlocal}
{G}(\theta_1,\theta_2)&=&\int_0^\infty dx \ x\left(x^2+\frac14\right) 
\ln \left| \frac{F(ix)}{F_a(ix)}\right| \nonumber\\
&&-2\int_0^\infty dx \ x\left(x^2+\frac14\right)
{\left(1+e^{2\pi x}\right)}^{-1}\ln \left|\pi F(ix) \right|,
\end{eqnarray}
the ``local'' functions $C_1(\theta)$ and $C_2(\theta)$ are
\begin{eqnarray}\label{eq:local}
C_1(\theta)&=&\int_0^\infty dx \ x^3 
            \left\{ \ln |\Sigma^S(x,\theta)|
             -\sum_{k=1}^{3}
	     s_k(\theta)x^{-k}-s_{4}(\theta)x^{-4}e^{-1/x}\right\}\nonumber \\
	     &&+\frac14\int_0^\infty dx \ x 
            \left\{ \ln |\Sigma^S(x,\theta)|
             -s_1(\theta)x^{-1}-s_{2}(\theta)x^{-2}e^{-1/x}\right\}+\frac12
	     s_4(\theta),\\
C_2(\theta)&=&\int_0^\infty dx \ x^3 
            \left\{ \ln |\Sigma^R(x,\theta)|
             -\sum_{k=1}^{3}
	     r_k(\theta)x^{-n}-r_{4}(\theta)x^{-4}e^{-1/x}\right\}\nonumber \\
	     &&+\frac14\int_0^\infty dx \ x 
            \left\{ \ln |\Sigma^R(x,\theta)|
             -r_1(\theta)x^{-1}-r_{2}(\theta)x^{-2}e^{-1/x}\right\}+\frac12
	     r_4(\theta),
\end{eqnarray}
and the constant part is
\begin{equation}\label{eq:const}
q=\chi \gamma \frac{17}{960} + \int_0^\infty dx \ x\left(x^2+\frac14\right) \ln
\left|\frac{\sinh\pi x}{\cosh \pi(x-i\mu)}\right|.
\end{equation}

The one-loop effective action in the conformally rescaled metric is
\begin{equation}
W_\Omega=-\frac12 \zeta'(0)-\frac12\zeta(0)\ln \mu_R^2,
\end{equation}
so it can be seen that the terms proportional to $\zeta(0)$ in Eq. (\ref{eq:zetaprime0})
can be absorbed into a redefinition of the renormalization scale $\mu_R$.

As a check on our results, the renormalization scale dependent term $\zeta(0)$ 
can also be calculated directly
using heat kernel methods. This is done in Section~\ref{sec:cocycle}. 
\section{Cocycle function}\label{sec:cocycle}

The cocycle function for the conformal rescaling was given in Eq. (\ref{eq:cocycle})
in terms of the $B_{5/2}(f,\Delta)$ heat kernel coefficient. 
The $B_{5/2}(f,\Delta)$ heat kernel coefficient 
has been calculated for general operators of Laplace type with mixed
Dirichlet and Robin boundary conditions in \cite{Branson:1999jz}. It is
comprised of curvature terms of order $R^2$ evaluated only on the boundary
of the spacetime. The general expression is quite lengthy. However,  
for a scalar field 
obeying Dirichlet boundary conditions, many terms are zero. Additionally,
the heat kernel coefficients simplify  further for the case of
a maximally symmetric boundary.

The calculation of the heat kernel coefficient is straightforward but tedious.
For the metric (\ref{eq:metric}) and operator (\ref{eq:operator}), we find
that the cocycle function can be written as
\begin{equation}
C[\Omega]=\sum_{i=1,2}\frac{1}{a^4(4\pi)^2}\int {|g^{(4)}|}^{1/2} d^4 x
\left\{\omega(z_i){\cal A}_i + {\cal B}_i\right\},
\end{equation}
where
\begin{equation}\label{eq:calA}
{\cal A}_i= \frac{17}{1920} + 
\frac{3}{16}\frac{1+\cos^2\theta_i}{\sin^4\theta_i}\left(\mu^2-1/4\right) 
- \frac{1}{8} \frac{1}{\sin^4\theta_i}\left(\mu^2-1/4\right)^2,
\end{equation}
and
\begin{equation}\label{eq:calB}
{\cal B}_i=-\frac{35}{192}-\frac{889}{1024}\cot^4\theta_i
-\frac{265}{256}\cot^2\theta_i-\frac{1}{64}\frac{\mu^2-4}{\sin^4\theta_i}
\left(17\cos^2 \theta_i+4\right),
\end{equation}
where we have reintroduced $\mu=\sqrt{4-20\xi+m^2/\sigma^2}$.

One can also use the heat kernel to directly evaluate $\zeta(0)$ through the
relation 
\begin{equation}
\zeta(0)=B_{5/2}(1,\Delta)=\sum_{i=1,2}\frac{1}{a^4(4\pi)^2}\int {|g^{(4)}|}^{1/2} d^4 x
\ {\cal A}_i.
\end{equation}
Inserting the expressions for $r_2(\theta)$ and $r_4(\theta)$ from Appendix~\ref{sec:coeff} into
the
expression for $\zeta(0)$ for the conformally transformed operator
(\ref{eq:zeta0}), it is easily seen that there is exact agreement between
the two calculations. These expressions also reduce to previously calculated values for flat branes
\cite{Garriga:2001ar, Moss:2004un} in the limit $a\rightarrow \infty$.

It is interesting to notice that the heat kernel coefficient
$B_{5/2}$ is comprised entirely of local geometrical objects, while the
full $\zeta$ function is a nonlocal quantity as information from both boundaries
is required for the eigenvalue problem. It is therefore quite remarkable that
$\zeta(0)$ can be obtained from the heat kernel coefficient, and this is a powerful check on
our method.
\section{Massless Conformally coupled case}
The integrals in equations (\ref{eq:nonlocal}) and (\ref{eq:local}) cannot be
done analytically in the general case we must resort to numerical methods. However, 
in the case $m=0$, and $\xi=3/16$, we find $\mu=1/2$ and the Legendre functions
simplify to hyperbolic functions and exponentials, so that
\begin{equation}
R^{-1/2}_{-1/2+ix}(\theta)= \frac{1}{x}\sqrt{\frac{2}{\pi}}
\sinh(x \theta),
\end{equation}
and
\begin{equation}
S^{1/2}_{-1/2+ix}(\theta)= \sqrt{\frac{2}{\pi}} e^{-x\theta}.
\end{equation}

All the coefficients $r_k(\theta)$ and $s_k(\theta)$ vanish. The local
functions
$C_1(\theta)$ and $C_2(\theta)$ also vanish. The constant $q$ can be absorbed into
a redefinition of the renormalization scale $\mu_R$.
The expression for $\zeta(0)$ also becomes simple. We find
\begin{equation}
\zeta(0)=\frac{1}{16\pi^2 a^4} \int |g^{(4)}|^{1/2} d^4 x \ \frac{17}{960}.
\end{equation}

The integral in the first term for the
nonlocal part $G(\theta_1,\theta_2)$ can be done explicitly in terms of the
Riemann zeta function $\zeta_R(s)$. The remaining integral in $G(\theta_1,\theta_2)$ must be none
numerically. We find
\begin{eqnarray}\label{eq:conscalar}
V_\Omega &= \displaystyle \frac{1}{16\pi^2 a^4}
&\left\{ -\frac{3\zeta_R(5)}{8 L^4} - \frac{\zeta_R(3)}{16 L^2} 
- 2\int_0^\infty dx \frac{x \left(x^2 + \frac14\right)}{e^{2\pi x}+1} \ln \frac{2 \sinh{L x}}{x}
 \right. \nonumber \\
 &&\left.-\frac{17}{1920}\ln \left(a^2 {\mu_R}^2\right)\right\},
\end{eqnarray}
where $L=\theta_2-\theta_1=z_2/a-z_1/a$.
From the previous equation, it is clear that the effective potential 
in the conformally rescaled metric is 
a function solely of the nondimensional conformal distance
$L$.

\begin{figure}[ht]
\begin{center}
\includegraphics[width=5in]{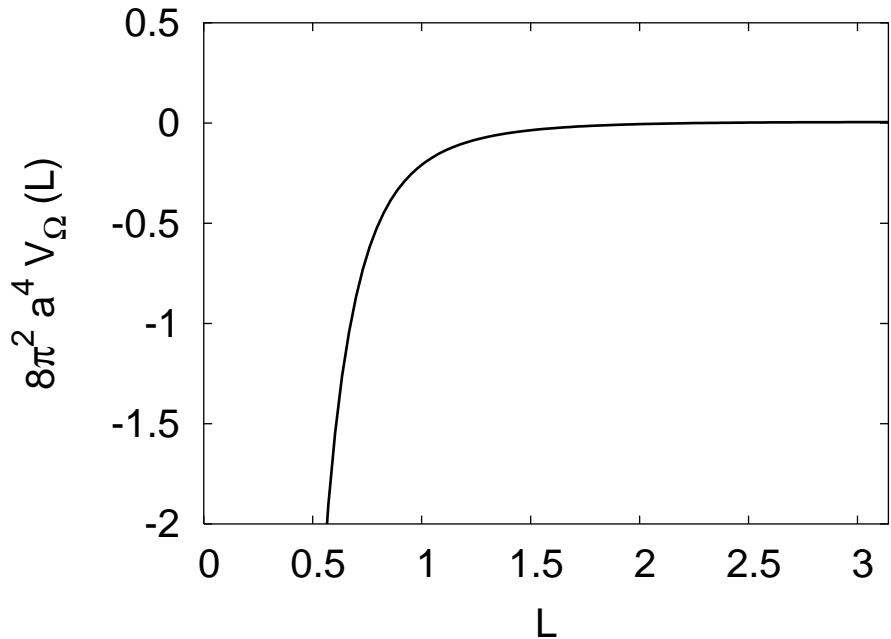}
\caption{Numerical results for the effective potential due to a conformally coupled scalar field as a function
of the non-dimensional conformal length $L=\theta_2-\theta_1=z_2/a-z_1/a$, where $z_1$ and $z_2$ are the positions of the
branes in conformal coordinate and $a$ is the $AdS_4$ radius of the brane at $z_1$}
\label{fig:conformal}
\end{center}
\end{figure}

We plot the effective potential $V_\Omega$ as a function of $L$ in 
Figure~\ref{fig:conformal}. We have adjusted the renormalisation scale so that
the maximum of the potential is at $V_\Omega=0$. 

To this result, we must also add the cocycle function, discussed in 
Section~\ref{sec:cocycle}. The total effective potential is then
\begin{equation}
V=V_\Omega+\sum_{i=1,2}\frac{1}{16\pi^2 a^4} \left\{\frac{17}{1920} \ln \sigma a \sin \theta_i
+\frac{5}{96}+\frac{25}{128}
\cot^2\theta_i+\frac{131}{1024}\cot^4\theta_i\right\}.
\end{equation}

The total effective potential $V$ is now no longer a function of $L=\theta_1-\theta_2$, but now
depends explicitly on $\theta_1$ and $\theta_2$.
The terms from the cocycle function can dominate the effective potential if 
$\theta_{1}$ or $\theta_2$ is small or close to $\pi$. Also, the terms due to the cocycle function
cannot be absorbed into a redefinition of the brane tensions as is often done
in the case of flat branes. 

It is worthwhile mentioning here that one cannot obtain
the effective action for a conformally coupled scalar field
in the background with de Sitter branes, obtained in
\cite{Elizalde:2002dd,Moss:2003zk},
by a simple
analytic continuation of $a\rightarrow ia$. This is because of the different global properties of
$S^4$ and $\mathrm{H}^4$, as has been noted in previous calculations of $\zeta$ functions
on $S^4$ and $\mathrm{H}^4$
\cite{Camporesi:1991nw}.
\section{More general cases}\label{sec:example}
\subsection{Numerical Results}\label{sec:numerical}
In the case of general mass and coupling constant $\xi$, the integrals in equations (\ref{eq:nonlocal}), (\ref{eq:local}) and
(\ref{eq:const}) must be computed numerically. As an example, we have chosen $m=0$ and $\xi=3/20$, so that $\mu=1$.
The integrals that must be evaluated are fairly computationally intensive, but can be performed simply (if slowly)
using computer packages such as MAPLE or MATHEMATICA.

\begin{figure}[ht]
\begin{center}
\includegraphics[width=5in]{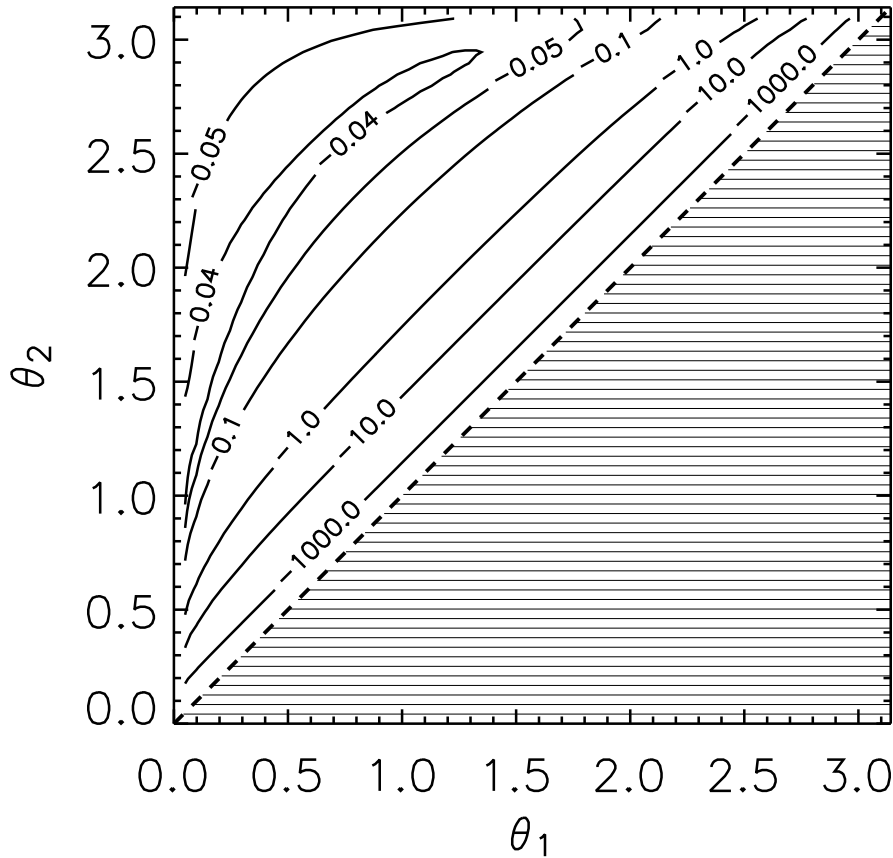}
\caption{Numerical results for the non-local part of the effective action. 
$\theta_1$ and $\theta_2$ are $z_1/a$ and $z_2/a$ respectively, where $z_1$ and
$z_2$ are the positions of the branes in the conformal coordinate $z$, and $a$ is the 
$AdS_4$ radius of the brane at $z_1$. 
The contours are values of $G(\theta_1,\theta_2)$. The
shaded region is disallowed, since we have restricted $z_2>z_1$}
\label{fig:nonlocal}
\end{center}
\end{figure}

The numerical results for the nonlocal part of the effective potential $G(\theta_1,\theta_2$ are shown as a 
function of $\theta_1=z_1/a$ and $\theta_2=z_2/a$ in
Figure~\ref{fig:nonlocal}. It can be seen from this figure that this
part of the effective action is not solely dependent on the difference
$\theta_2-\theta_1$, unlike the conformally coupled case considered above.  
Also, this term in the effective action diverges strongly to negative values
as $\theta_2-\theta_1\rightarrow 0$. 

\begin{figure}[ht]
\begin{center}
\includegraphics[width=5in]{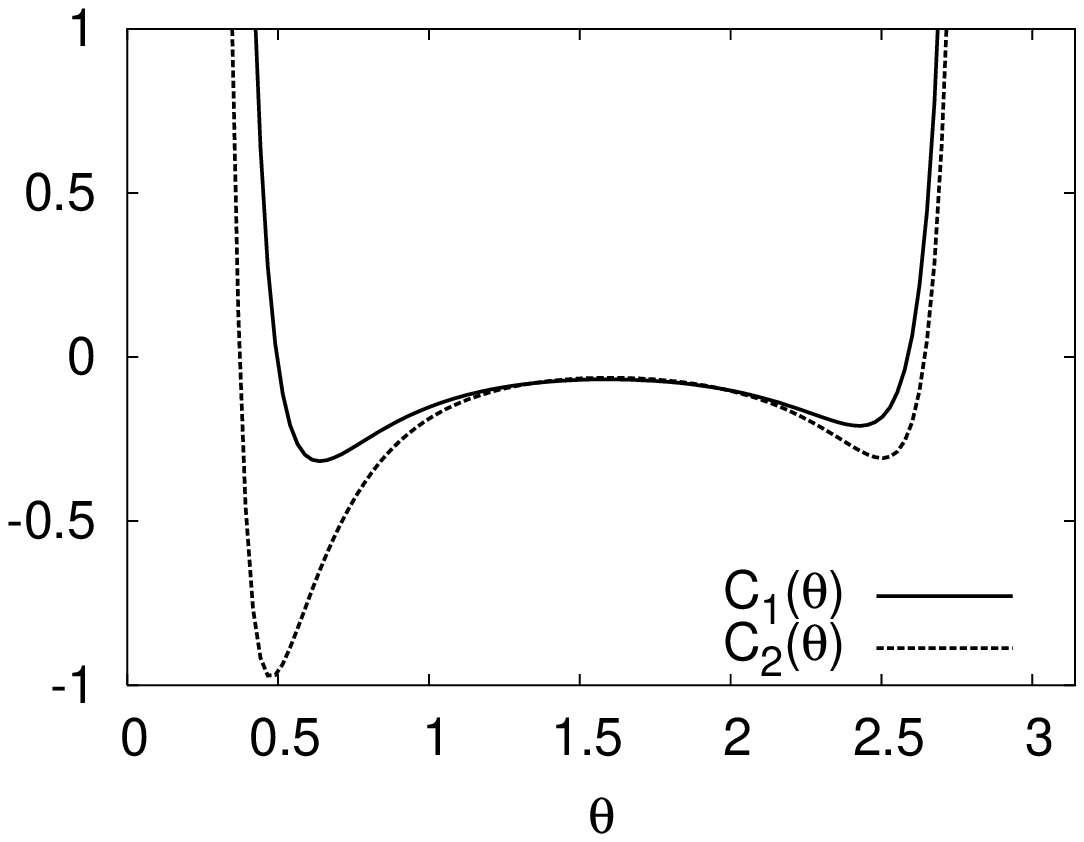}
\caption{Numerical results for the local functions $C_1(\theta)$ and 
$C_2(\theta)$ in the effective potential.
Again, $\theta_1=z_1/a$ and $\theta_2=z_2/a$ where $z_1$ and $z_2$ are the
positions of the branes in conformal coordinates, and $a$ is the $AdS_4$ radius of
the brane at $z_1$.}
\label{fig:local}
\end{center}
\end{figure}

The local functions $C_1(\theta)$ and $C_2(\theta)$ are shown in Figure~\ref{fig:local}. 
These terms can  become
significant if $\theta_{1}$ or $\theta_2$ are small or close to $\pi$. Finally, We calculate the
constant $q$ to be
\begin{equation}
q\approx -0.03128. 
\end{equation}

These numerical results enable one to calculate the finite part of the effective action
of the conformally rescaled operator. However, to this, we must also add the
renormalization
scale dependent term $\zeta(0)\ln \mu_R^2$.
From both Eq. (\ref{eq:zeta0}) and the $B_{5/2}(1,\Delta)$ heat kernel 
coefficient, one can calculate the value of $\zeta(0)$. Using either method,
we find
\begin{equation}
\zeta(0)=\frac{1}{16\pi^2 a^4}\sum_{i=1,2} \int |g^{(4)}|^{1/2} d^4 x \left\{
\frac{19}{240}+\frac{9}{32}\cot^2\theta_i+\frac{27}{128}\cot^4\theta_i\right\}.
\end{equation}

We have now calculated all the terms in the effective potential in the conformally
rescaled metric $V_\Omega$.
The total effective potential including the cocycle function is then
\begin{eqnarray}
V=V_\Omega+\sum_{i=1,2}\frac{1}{16\pi^2 a^4} &&
\left\{\left(\frac{19}{240}+\frac{9}{32}\cot^2\theta_i+\frac{27}{128}\cot^4\theta_i\right)
\ln \sigma a \sin \theta_i \right. \nonumber \\
&&\left.+\frac{1}{192}-\frac{13}{256}
\cot^2\theta_i-\frac{73}{1024}\cot^4\theta_i\right\}.
\end{eqnarray}

\subsection{Analytical approximation} 
The numerical results in Section~\ref{sec:numerical} show that the effective potential becomes very large 
if $L=\theta_2 -\theta_2\ll 1$. In this case, the dominating contribution to the effective potential is from the
integral in the first term in Eq. (\ref{eq:nonlocal}). We can approximate this
integral using the asymptotic expansion of the Legendre functions. This gives a series in powers of
$L$. The first two terms in this expansion of the effective potential are 
\begin{equation}\label{eq:approx}
V \approx -\frac{3\zeta_R(5)}{128\pi^2(z_2-z_1)^4}
+\frac{\zeta_R(3)}{64\pi^2(z_2-z_1)^2}\left[\sigma^2 \left(\mu^2-1/4\right) -\frac{1}{4a^2}\right] + O(\ln(z_2-z_1).
\end{equation}
The first term on the right hand side of Eq. (\ref{eq:approx}) is the Casimir potential for two flat branes in
flat space, separated by a distance $z_2-z_1$. Note that this leading term is independent of the
mass or coupling to the curvature of the scalar field. 
A similar result was found for the small distance approximation with flat branes in
\cite{Garriga:2001ar}.

The second term in Eq. (\ref{eq:approx}) is the first order correction to this result due to
bulk and brane curvature. This approximation results in a much more manageable expression than the exact expression
for the effective potential.


\section{Conclusions}

We have calculated the one-loop effective potential for a scalar field 
with general mass and coupling to the Ricci scalar in
the two brane Randall-Sundrum model with detuned brane tensions such that the boundary
branes are $AdS_4$. Conformal rescalings of the metric are used to relate the
metric to two hyperbolic branes. In general, the resulting expressions contain integrals
of Legendre functions which must be performed numerically. We obtain some
approximations in the small conformal distance limit, which reduce to the 
Casimir potential for two flat branes in flat space, and also calculate the first
order corrections due to the brane curvature. Our results are
checked by comparing the renormalization scale dependence obtained from the
conformally rescaled operator with a direct computation using heat kernel
coefficients. They are found to agree exactly.

It should be remembered that, in the dimensionally reduced theory, there is
also a classical potential for the radion field, unless the brane tensions are
tuned. It is interesting to compare the form of the classical
radion potential and the quantum effective potential for the radion due
to a bulk scalar field. The classical radion effective potential has been analyzed
in \cite{Bagger:2003dy}. The classical potential is most naturally
expressed in terms of the proper distance $r$, related to $z_1$ and $z_2$ by
\begin{equation}
\pi r=\int_{z1}^{z2} e^{-w(z)} dz = \frac{1}{\sigma} \ln \tan \frac{z_2}{2a}-\frac{1}{\sigma} \ln \tan \frac{z_1}{2a}
\end{equation} 
For small conformal separation, $\pi r \approx  z_2-z_1$
and the classical effective potential goes like
\begin{equation}
V_\text{classical} \approx \text{const} \ (\sigma a \pi r)^{-2} \approx \text{const} \frac{1}{(\sigma a(z_2-z_1))^2}
\end{equation}
where ``$\text{const}$''  is a positive constant. One can see from Eq. (\ref{eq:approx})
that the effective potential generated by a quantized bulk scalar field would
destabilize the
classical potential at small separation, as the quantum effective potential diverges faster
than the classical potential in this limit. 
Of course, there will also be an effective
potential generated by graviton fluctuations which have not been included in this discussion. 

It would be
interesting to extend this calculation to higher spin fields so that one could investigate the
effect of the supersymmetry breaking on the Casimir potential. An intriguing possibility in the 
supersymmetric Randall-Sundrum model is that one-loop effects could generate a potential for the 
supersymmetry breaking parameter (either the
twist angle of the gravitino boundary condition or the v.e.v. of the fifth component of the gauge field) 
which is a modulus of compactification at the
classical level \cite{Bagger:2003fy}. This
situation is reminiscent of supersymmetry breaking in heterotic M theory by gaugino condensation \cite{Horava:1996vs}. 

\acknowledgments
I thank Ian Moss for helpful discussions and comments on an earlier draft of this paper.
\appendix 
\section{Robin boundary conditions}\label{sec:robin}
\subsection{$\zeta$-function on $\mathrm{H}^4 \times I$}
Robin boundary conditions have some combination of the field and its normal derivative
vanishing on the boundary. We take
\begin{equation}
\left(\partial_N + \eta K \right) \phi=0 \quad \text{on} \quad \partial {\cal M}.
\end{equation}
where $K$ is the trace of the extrinsic curvature of the boundary and $N$ denotes the \emph{outward}
pointing unit normal.
Under a conformal transformation, this changes to
\begin{equation}
\left(\partial_N + \hat{\eta} K \Omega^{-1} \right) \tilde{\phi} =0\quad \text{on} 
\quad \partial {\cal M},
\end{equation}
where $\hat{\eta}=\eta-3/8$ since the field $\phi$ also transforms under the conformal rescaling.
Applying this boundary condition to the general solution, 
the implicit equation for the $b_n$'s can be given in terms of new functions $T^{-\mu}_{\nu}(\theta)$ and 
$U^{\mu}_{\nu}(\theta)$, defined by
\begin{subequations}
\begin{eqnarray}
T^{-\mu}_{\nu}(\theta)&=&R^{-\mu+1}_\nu(\theta)+\cot \theta
\left(1/2-\mu-4\hat{\eta} \right) R^{-\mu}_\nu(\theta),  \\
U^{\mu}_{\nu}(\theta)&=&S^{\mu+1}_\nu(\theta)+\cot \theta \left(1/2+\mu-4\hat{\eta} \right) S^{\mu}_\nu(\theta),
\end{eqnarray}
\end{subequations}
as
\begin{equation}
F(b_n)=T^{-\mu}_{-1/2+b_n} (\theta_1)U^{\mu}_{-1/2+b_n}
(\theta_2)-T^{-\mu}_{-1/2+b_n} (\theta_2)U^{\mu}_{-1/2+b_n}(\theta_1)=0.
\end{equation}
The procedure then follows the Dirichlet case, in that we write the $\zeta$ function as
a contour integral, shift the contour to the real axis, and then analytically
continue to find the effective action.
If we define the functions $\Sigma^T(x,\theta)$ and $\Sigma^U(x,\theta)$ by
\begin{equation}
T^{-\mu}_{-1/2+ix}
(\theta) = e^{i\pi \mu/2+i\pi/4} \frac{x}{\sqrt{2\pi }} e^{x \theta}
\frac{\Gamma(ix-\mu+1/2)}{\Gamma(ix+1)} \Sigma^T(x,\theta),
\end{equation}
and
\begin{equation}
U^{\mu}_{-1/2+ix}
(\theta) = -e^{i\pi \mu/2 -i\pi/4}x\sqrt{\frac{2}{\pi}} e^{-x \theta}
\frac{\Gamma(ix+\mu+1/2)}{\Gamma(ix+1)} \Sigma^U(x,\theta),
\end{equation}
then $\Sigma^T(x,\theta)$ and $\Sigma^U(x,\theta)$ have asymptotic expansions of the Poincar\'e type.
Similar to the Dirichlet case, we define coefficients $t_k(\theta)$ and $u_k(\theta)$ by
\begin{equation}
\ln |\Sigma^T(x,\theta)|\sim\sum_{k=1}^\infty t_k (\theta) x^{-k}+O(e^{-x}),
\quad \ln |\Sigma^U(x,\theta)|\sim\sum_{k=1}^\infty u_k (\theta) x^{-k}.
\end{equation}
We again refer the reader to Appendix~\ref{sec:coeff} for explicit expressions for
$t_k(\theta)$ and $u_k(\theta)$.

Thus, for large $x$,
\begin{equation}
\ln |F(ix)| \sim  (\theta_2-\theta_1)x +\chi \ln(x) -\ln \pi + O(x^{-2}),
\end{equation}
where now
\begin{equation}
\chi=1.
\end{equation}
The effective action in the conformally rescaled metric for Robin boundary 
conditions can now be
found by substituting $R$, $S$, $\Sigma^R$, $\Sigma^S$, $r_k$ and $s_k$ in the 
Dirichlet case by $T$, $U$, $\Sigma^T$, $\Sigma^U$, $t_k$ and
$u_k$. Additionally, one must replace $\chi=-1$ for Dirichlet boundary conditions by
$\chi=+1$ for Robin boundary conditions.

One subtlety in the Robin case is that there may exist zero or imaginary solutions
of $F(b_n)=0$. For example, a conformally coupled scalar field with Robin boundary
conditions has a zero mode. This means that there will be a contribution to the $\zeta$-function
from integral over the 
small semi-circle ${\cal C}_\varepsilon$ as $\varepsilon\rightarrow 0$. Imaginary values
of $b_n$ signal an instability, and these situations should be physically unacceptable. In our analysis, 
we assume that the values of $\eta$, $\xi$ and $m$ are such that there are no
zero or imaginary solutions of $F(b_n)=0$.

\subsection{Cocycle function}

The heat kernel coefficient for scalar fields obeying Robin boundary conditions is
a little more lengthy than the Dirichlet case. Again, the calculation is straightforward, but
messy.
For the cocycle function, we find that Eqs. (\ref{eq:calA}) and
(\ref{eq:calB}) become
\begin{eqnarray}
{\cal A}_i&=&-\frac{17}{1920} + 64 \cot^4\theta_i \hat{\eta}^4
+2\cot^2\theta_i\hat{\eta}^2
+\frac{1}{8}\frac{1}{\sin^4\theta_i}(\mu^2-1/4)^2 \nonumber \\
&&-\frac{1}{\sin^4\theta_i}\left[\cos^2\theta_i(8\hat{\eta}^2+2\hat{\eta})
+\frac{3}{16}(\cos^2\theta_i+1)\right](\mu^2-1/4),
\end{eqnarray}
and
\begin{eqnarray}
{\cal B}_i&=&\left(\frac{61}{15360}+\frac{11}{48}\hat{\eta}
-\frac23\hat{\eta}^2-8\hat{\eta}^3\right)\cot^4\theta_i
+\left(\frac{623}{768}+\frac{71}{24}\hat{\eta}
-\hat{\eta}^2\right)\frac{\cos^2\theta_i}{\sin^4\theta_i}
\nonumber \\
&&+\frac{35}{192}\frac{1}{\sin^4\theta_i}+\left[\left(\hat{\eta}+\frac{15}{64}\right){\cos^2\theta_i}
+\frac{1}{16}\right]\frac{\mu^2-4}{\sin^4\theta_i}.
\end{eqnarray}
respectively.

Again, these results reduce to the previously known flat brane values calculated in \cite{Moss:2004un}
in the limit $a\rightarrow\infty$.

\subsection{An example}
In some special cases, the implicit equation for the eigenvalues for 
Robin boundary conditions 
can be reduced to the one resembling that for a Dirichlet
case, plus one ``extra'' mode. 
As an example we consider the case where $\mu=0$ and $\eta=1/2$. Then the implicit equation
for the $b_n$ reduces to
\begin{equation}
\left(b_n^2-\frac14\right)\left(R^{-1}_{-1/2+b_n} (\theta_1)S^{1}_{-1/2+b_n}
(\theta_2)-R^{-1}_{-1/2+b_n}
(\theta_2)S^{1}_{-1/2+b_n}(\theta_1)\right)
=0.
\end{equation}
This is simply the example considered in Section~\ref{sec:example}, but with the ``extra'' mode
at $b_n=1/2$.
We can add this extra mode to the $\zeta$-function considered in Section~\ref{sec:example} by hand.
This gives an extra contribution to $\zeta(0)$ of $-(240\pi^2 a^4)^{-1}$, giving
\begin{equation}
\zeta(0)=\frac{1}{16\pi^2 a^4}\sum_{i=1,2} \int |g^{(4)}|^{1/2} d^4 x \left\{
\frac{11}{240}+\frac{9}{32}\cot^2\theta_i+\frac{27}{128}\cot^4\theta_i\right\},
\end{equation}
again in agreement with the value calculated directly from the heat kernel
coefficient $B_{5/2}(1,\Delta)$.
The extra mode will also give an extra contribution
to $\zeta'(0)$ of a constant (independent of $\theta_1$ or $\theta_2$).

However, the cocycle function cannot be deduced from the Dirichlet case, and
must be computed explicitly.

\section{Coefficients in asymptotic expansions of the Legendre functions}\label{sec:coeff}
\subsection{Dirichlet boundary conditions}\label{sec:dirchletcoeff}
From the asymptotic expansion of the Legendre functions for large degree \cite[page 146]{Bateman}, the asymptotic expansion of 
$\ln |\Sigma^R(x,\theta)|$ and $\ln |\Sigma^S(x,\theta)|$ can be shown to be of the
form
\begin{equation}
\ln |\Sigma^R(x,\theta)|\sim\sum_{k=1}^\infty r_k (\theta) x^{-k},
\quad \ln |\Sigma^S(x,\theta)|\sim\sum_{k=1}^\infty s_k (\theta) x^{-k}.
\end{equation}
The coefficients are easily evaluated with the help of a computer algebra package such as MAPLE
or MATHEMATICA.
We obtain the first four coefficients as
\begin{subequations}
\begin{eqnarray}
r_1(\theta) & = & -\frac12 \cot \theta \left(\mu^2-1/4\right), \\
r_2(\theta) & = & -\frac{1}{4} \frac{1}{\sin^2\theta}\left(\mu^2-1/4\right), \\
r_3(\theta) & = & -\frac{1}{96} \frac{\cos \theta}{\sin^3 \theta} \left(\mu^2-1/4\right)
\left(8\mu^2\cos^2 \theta -12\mu^2 - 2\cos^2 \theta + 27 \right),\\
r_4(\theta) & = & \frac{1}{32}\frac{1}{\sin^4\theta}\left(\mu^2-1/4\right)
\left(4\mu^2-8\cos^2\theta-5\right),
\end{eqnarray}
\end{subequations}
and $s_n(\theta)=(-1)^n r_n(\theta)$.

\subsection{Robin boundary conditions}\label{sec:robincoeff}
Similarly to the Dirichlet case, for Robin boundary conditions we have 
\begin{equation}
\ln |\Sigma^T(x,\theta)|\sim\sum_{k=1}^\infty t_k (\theta) x^{-k},
\quad \ln |\Sigma^U(x,\theta)|\sim\sum_{k=1}^\infty u_k (\theta) x^{-k}.
\end{equation}
This time, we find
\begin{subequations}
\begin{eqnarray}
t_1(\theta) & = & -\frac{1}{2}\cot\theta \left(\mu^2-1/4+8\hat{\eta}\right),\\
t_2(\theta) & = & \frac{1}{4\sin^2\theta}\left(\mu^2-1/4-32\hat{\eta}^2\cos^2\theta
\right),\\
t_3(\theta) & = &  -\frac{1}{384}\frac{\cos\theta}{\sin^3\theta}\left(
21-72\mu^2-48\mu^4+2\cos^2\theta-16\mu^2\cos^2\theta\right. \nonumber \\
&&\quad \quad \quad \quad \left.+32\mu^4\cos^2\theta
+192\hat{\eta}+8192\hat{\eta}^3\cos^2\theta-768\mu^2\hat{\eta} \right),\\
t_4(\theta) & = &-\frac{1}{128\sin^4\theta}
\left(5-24\mu^2+16\mu^4+8\cos^2\theta+8192\hat{\eta^4}\cos^4\theta 
-32\mu^2\cos^2\theta
\right. \nonumber \\
&&\quad \quad \quad \quad
\left.+256\hat{\eta}^2\cos^2\theta+64\hat{\eta}\cos^2\theta-256\mu^2\hat{\eta}\cos^2\theta-1024\mu^2\hat{\eta}^2\cos^2\theta\right),
\end{eqnarray}
\end{subequations}
and $u_i(\theta)=(-1)^i t_i(\theta)$.


\bibliography{ads}
\end{document}